\documentstyle[aps,prb,twocolumn]{revtex}
\begin{document}
\draft
\title{Neutron Scattering Study of Crystal Field Energy Levels and Field
Dependence
of the Magnetic Order in Superconducting HoNi$_{{\bf {2}}}$B$_{{\bf {2}}}$C}
\author{T. E. Grigereit$^{1,2}$ and J. W. Lynn$^{1,2}$}
\address{$^{1}$Reactor Radiation Division, National Institute of \\
Standards and Technology,\\
Gaithersburg, Maryland 20899}
\address{$^{2}$Center for Superconductivity Research,\\
University of Maryland,\\
College Park, Maryland 20742}
\author{R. J. Cava, J. J. Krajewski, and W. F. Peck, Jr.}
\address{AT\&T Bell Laboratories, Murray Hill, NJ 07974}
\maketitle

\begin{abstract}
Elastic and inelastic neutron scattering measurements have been carried out
to investigate the magnetic properties of superconducting (${\text{T}}_c\sim
8$~K) HoNi$_{{\bf {2}}}$B$_{{\bf {2}}}$C. The inelastic measurements reveal
that the lowest two crystal field transitions out of the ground state occur
at 11.28(3) and 16.00(2)~meV, while the transition of 4.70(9) meV between
these two levels is observed at elevated temperatures. The temperature
dependence of the intensities of these transitions is consistent with both
the ground state and these higher levels being magnetic doublets. The system
becomes magnetically long range ordered below $8K$, and since this ordering
energy $kT_N\approx 0.69meV$ $\ll 11.28meV$ the magnetic properties in the
ordered phase are dominated by the ground-state spin dynamics only. The low
temperature structure, which coexists with superconductivity, consists of
ferromagnetic sheets of Ho$^{3+}$ moments in the {\em a-b} plane, with the
sheets coupled antiferromagnetically along the {\em c}-axis. The magnetic
state that initially forms on cooling, however, is dominated by an
incommensurate spiral antiferromagnetic state along the {\em c}-axis, with
wave vector ${\text{q}}_{\text{c}}\sim 0.054~{\text{\AA }}^{-1}$, in which
these ferromagnetic sheets are canted from their low temperature
antiparallel configuration by $\sim 17^{\circ }$. The intensity for this
spiral state reaches a maximum near the reentrant superconducting transition
at $\sim $5~K; the spiral state then collapses at lower temperature in favor
of the commensurate antiferromagnetic state. We have investigated the field
dependence of the magnetic order at and above this reentrant superconducting
transition. Initially the field rotates the powder particles to align the
{\em a-b} plane along the field direction, demonstrating that the moments
strongly prefer to lie within this plane due to the crystal field
anisotropy. Upon subsequently increasing the field at constant T the
antiferromagnetic and spiral states are both observed to decrease in
intensity, but at modest fields the spiral state decreases much less
rapidly. Approaching the superconducting phase boundary from high fields, we
find that the spiral state is strongly preferred, in deference to the
superconductivity, again demonstrating a direct coupling between these two
cooperative phenomena. The magnitude of the spiral wave vector q$_{\text{c}}$%
, on the other hand, shows very little field dependence. A magnetic moment
of 8.2$\pm 0.2$~$\mu _{\text{B}}$ for the Ho$^{3+}$ is obtained from the
observed field dependence of the induced moment at high fields ($7T$).
\end{abstract}

\pacs{PACS numbers: 75.25.+z, 74.70.Ad, 75.30.Kz}

\newpage\

\narrowtext

\section{Introduction}

The new classes of quaternary intermetallic superconductors, namely the
borocarbide series\cite{CavaYSC,NagGupDisc,CavaRESC,SiegCavaStr} RNi$_2$B$_2$%
C, where R is a rare-earth ion, and more recently the boronitride series\cite
{Cavanitride,Mattnitride} R$_3$Ni$_2$B$_2$N$_3$, exhibit relatively high
transition temperatures (up to 23~K for YPd$_2$B$_2$C)\cite{CavaYSC} while
possessing the possibility of long-range magnetic order of the rare-earth
subsystem over the same temperature range. In particular, the interplay
between superconductivity and magnetism, previously limited to the ternary
Chevrel-phase and related systems,\cite{FischerMaple} is realized in a
dramatic fashion in these new materials, with the most interesting system
being HoNi$_2$B$_2$C. This material becomes superconducting\cite{Eisaki} at $%
T_C\approx 8K$ while developing incommensurate long range magnetic order at
about the same temperature.\cite{Grigereit} At $\sim 5K$ the
superconductivity is reentrant as evidenced by a deep minimum in H$_{\text{c}%
2}$ near 5~K, below which the incommensurate magnetic order is suppressed in
favor of a simple commensurate antiferromagnetic structure, which allows the
return of superconductivity and a coexistence at low T.

Following the initial discovery\cite{CavaYSC,NagGupDisc,CavaRESC,SiegCavaStr}
of these materials, and the bulk measurements by Eisaki {\em et al.} \cite
{Eisaki} that revealed a competition between magnetism and superconductivity
in HoNi$_2$B$_2$C, we reported\cite{Grigereit} the first magnetic
diffraction data and solution of the magnetic structures in this material.
We also carried out complete profile refinements of both the nuclear and
magnetic structures.\cite{Huang} Fig.~1(a) shows the body-centered
tetragonal unit cell (I4/mmm), which consists of Ho-C planes separated by Ni$%
_2$B$_2$ layers stacked along the {\em c}-axis. The lattice parameters at
room temperature are $a=3.5170(1)~{\text{\AA }}$ and $c=10.5217(3)~{\text{%
\AA }}$ as obtained from the profile refinement. The Ho moments order at $%
\sim 8K$, with two types of magnetic ordering being observed. The first is
shown in Fig. 1(b), where we have ferromagnetic sheets of spins in the {\em %
a-b} plane which are coupled antiferromagnetically along the {\em c}-axis.
However, satellite peaks are also observed along the {\em c}-axis and
correspond to the direction of the spins in each sheet being rotated $\sim
16.6^{\circ }$ away from the antiferromagnet compensation. Thus there is an
antiferromagnet spiral that forms along the {\em c}-axis. If this angle were
exactly 15$^{\circ }$ as suggested in the figure then this would be a long
wavelength commensurate spiral, with a period of 24 Ho layers (12 unit
cells). However, the actual spiral that forms is incommensurate with the
underlying lattice.

Intensity for both types of peaks are observed below $\sim 8K$, and they
increase with decreasing temperature. However, we have found that the
relative intensities are different for different samples, and thus we
believe that these two magnetic phases are coming from separate regions of
the samples; further work on these aspects, including studies of doped
systems, will be reported elsewhere.\cite{Lynn} The intensities for both
magnetic phases increase with decreasing temperature, but in the vicinity of
the reentrant superconducting transition at $\sim 5K$ the intensity of the
spiral state suddenly begins to rapidly decrease while the intensity for the
commensurate antiferromagnetic peak rapidly increases, until it saturates at
low temperatures. This low temperature commensurate antiferromagnetic state
coexists with superconductivity.

The behavior of HoNi$_2$B$_2$C contrasts sharply with that for ErNi$_2$B$_2$%
C,\cite{Zarestky,SinhaEr} the only other Ni-containing
magnetic-superconductor system that has been investigated so far with
neutrons. Here an {\em a}-axis spin density wave state is observed at all
temperatures, and this state readily coexists with superconductivity over
the full temperature range where these two cooperative states are observed.
Subsequent to our initial work on HoNi$_2$B$_2$C, Goldman et al.\cite
{GoldmanStassisAmes} reported similar data on single crystal samples, and
also found a small {\em a}-axis modulation\cite{TomyOakridge} above the
reentrant transition similar to the large {\em a}-axis peaks observed in the
Er material. We have also found a small {\em a}-axis modulation, but over a
temperature-shifted and narrower T range than what Goldman et al.\cite
{GoldmanStassisAmes} found; these results will be reported elsewhere.\cite
{Lynn} However, since only the Ho system exhibits a giant anomaly in H$_{c2}$
and reentrant superconductivity, and only the Ho system possesses the {\em c}%
-axis spiral, it is clear that these two phenomena are directly related to
each other.

\section{Experimental Configurations}

The HoNi$_2$B$_2$C polycrystalline sample was prepared by arc-melting and
subsequent annealing,\cite{CavaRESC} using the $^{11}$B isotope to reduce
nuclear absorption. Elastic neutron scattering was performed on the BT-2 and
BT-9 triple-axis spectrometers at the National Institute of Standards and
Technology (NIST) Research Reactor. A pyrolytic graphite monochromator was
used in each case, with a pyrolytic graphite filter to remove higher order
wavelength contaminations. For the diffraction measurements typical
collimations of 60$^{\prime }$-20$^{\prime }$-20$^{\prime }$ were employed,
with no analyzer crystal (double-axis mode), with a neutron wavelength of
2.35~\AA . For the field-dependent measurements a vertical-field 7T
superconducting magnet was utilized. For the inelastic measurements we used
a pyrolytic graphite (002) analyzer crystal to select a fixed scattered
neutron energy E$_F$ of 14.8 meV.

\section{Inelastic Neutron Scattering}

In order to understand the nature of the magnetic order and the ground-state
spin dynamics, it is important in these heavy rare earth systems to
determine the single-ion crystal-field splittings of the Ho$^{3+}$ ions. We
therefore carried out inelastic neutron scattering measurements to
investigate the low-lying crystalline electric field (CEF) energy levels of
HoNi$_2$B$_2$C. For tetragonal D$_{\text{4}h}$ point group symmetry the
17-fold degenerate $^5{\text{I}}_8$ energy levels of Ho$^{3+}$ are split
into nine (non-magnetic) singlets and four (magnetic) doublets.\cite{Loong}
The heavy rare earths typically have CEF energies $\ll $ spin-orbit
J-splittings so that J-mixing may be neglected to a good approximation and
the ground state may be taken to be due to the CEF perturbation of the
lowest J-multiplet. As the holmium ions order magnetically we anticipate
that the crystal field ground state is very likely a doublet.

The scattering intensity is proportional to the neutron scattering
cross-section for crystal field transitions among N noninteracting ions,
which is given in the dipole approximation as\cite{Trammell,Lovesey}
\begin{equation}
I\sim \frac{d^2\sigma }{d\Omega dE}=N\frac{k_i}{k_f}\frac{\left( \gamma
r_0\right) ^2}4g_J^2S({\bf Q},E),  \label{Intensity}
\end{equation}
with the scattering function
\begin{eqnarray}
S({\bf Q},E) &=&f^2({\bf Q})e^{-2W({\bf Q})}\sum_{i,f}\frac{e^{-E_i/k_BT}}Z
\nonumber \\
&&\times |\langle f|J_{\perp }|i\rangle |^2\delta (E_f-E_i-E)
\end{eqnarray}
where $k_{i,f}$ are the incoming (scattered) wave vectors and $E_{i,f}$ the
initial (final) energies of the crystal field transitions, which are
independent of the momentum transfer ${\bf Q}$ since this is a single ion
effect, and $|i\rangle $ is the CEF state corresponding to the i$^{th}$
level within the J-multiplet, and the other symbols have their usual meaning.%
\cite{Lovesey} The matrix elements are between CEF states connected by the
component of the total angular momentum operator, $J_{\perp }$, which is
perpendicular to the momentum transfer $\hbar {\bf Q}$. The delta function
serves to conserve energy by equating the neutron energy transfer E with the
change in CEF level energy, E$_i$ - E$_f$. The intensity depends on ${\bf Q}$
through the Debye-Waller term, exp(-2W$_{{\bf Q}}$), and the magnetic form
factor f(${\bf Q}$), the latter of which is equivalent to the Fourier
transform of the magnetization density and is available for the rare-earths
in tabulated form\cite{formfactors}. For our case the Debye-Waller term is
small and the ${\bf Q}$-dependence is dominated by f(${\bf Q}$), causing the
intensity to decrease monotonically with increasing $Q$.

The temperature dependence arises through the occupational probabilities $P_{%
\text{i}}$, which are described by Boltzmann statistics for noninteracting
and distinguishable ions as
\begin{equation}
P_i=g_ie^{-E_i/k_BT}/Z
\end{equation}
where the partition function
\begin{equation}
Z=\sum_ie^{-E_i/k_BT}
\end{equation}
and g$_{{\rm i}}$ is the degeneracy of the i$^{{\rm th}}$ level. Neutron
energy-loss processes (E $>$ 0) dominate at low temperatures since the
Boltzmann probabilities are small for states above the ground state.

Equation \ref{Intensity} above may be simplified to
\begin{equation}
I\sim C\sum_i\frac{P_ig_i}Z  \label{Cratio}
\end{equation}
where all of the temperature-independent factors have been collected into a
single constant, {\em C}, except for the degeneracy of each level, g$_{\text{%
i}}$, and where the (small) Debye-Waller factor has been ignored. The ratio
of matrix elements for particular transitions may then be formed as the
ratio of their {\em C} values since all other terms cancel to a good
approximation.

Fig.~2 shows an example (at 6 K) of the observed inelastic spectra. We chose
to make these measurements at a momentum transfer of $Q\approx 2{\text{\AA }}%
^{-1}$, which avoids any Bragg peaks at the elastic position while keeping $%
Q $ small to obtain a strong magnetic signal and avoid significant phonon
scattering. The strong elastic scattering is due to the elastic crystal
field scattering plus nuclear incoherent scattering. This type of
measurement, taken at a series of temperatures, reveals two crystal field
excitations from the ground state to the first CF level at 11.28(3), and
from the ground state to the second level at 16.00(2) meV. We also observed
some additional quasielastic scattering below $\sim 2$ meV, which we
attribute to the spin dynamics within the crystal field ground state. Below
the ordering temperature of $\sim 8$ K ($\rightarrow 0.69$ meV) this is the
(powder-averaged) spin wave scattering, while at higher temperatures this is
just the paramagnetic diffuse scattering. We note that since the energetics
associated with the ordering is much less than the energy of the first
crystal field level (0.69 $\ll $ 11.28), below the ordering temperature only
the crystal field ground state is significantly populated. Hence the
symmetry of the magnetic order parameter and the spin dynamics in the
ordered state will be controlled by the properties of the crystal field
ground state.

At low temperature we can only observe excitations out of the ground state,
but at elevated temperature the levels at 11.28 and 16.00 meV will become
populated and we should then see an additional excitation at 4.72 meV. This
transition is indeed observed at 4.70(9) meV. Fig.~3 shows the observed
temperature dependence of the intensities of the two transitions out of the
ground state and the transition between them, each point of which is the
result of the fit with Gaussian lineshapes such as shown in Fig.~2. Over the
range of temperatures up to 150 K the energies of the two transitions out of
the ground state are temperature-independent within experimental error. The
temperature dependence of the scattering intensities of each transition has
been modeled with Boltzmann statistics, indicated by the dashed lines in
Fig.~3. It is clear that the CF ground state is a magnetic doublet, and
indeed all three levels were found to be best fit by doublets. The 11.28 and
16.00 meV transitions follow the thermal population of the ground state,
which decreases with increasing temperature as the excited levels are
thermally populated, while the intensity of the 4.7 meV transition increases
with temperature and appears to saturate near 150~K (the highest temperature
measured). The matrix element ratios are found to be
\begin{equation}
\frac{|\langle 0|J_{\perp }|11.28\rangle |^2}{|\langle 0|J_{\perp
}|16.00\rangle |^2}=3.04
\end{equation}
and
\begin{equation}
\frac{|\langle 0|J_{\perp }|11.28\rangle |^2}{|\langle 11.28|J_{\perp
}|16.00\rangle |^2}=0.99
\end{equation}
using equation \ref{Cratio}. The elastic magnetic scattering intensity was
also found to decrease with increasing temperature.

The field-dependent data we present below reveal that the moments strongly
prefer to lie in the {\em a-b} plane, while the crystal field data
demonstrate that the ground state is doubly degenerate. The exchange
interactions are three dimensional, but they are clearly different along the
{\em c}-axis then in the {\em a-b} plane. Thus we expect that the magnetic
Hamiltonian will correspond to a $S=\frac 12$, three dimensional {\em xy}
system. It is also likely that a four-fold in-plane anisotropy will be
needed to adequately describe the system. The indirect (RKKY) exchange
interactions, on the other hand, must be relatively complicated as evidenced
by the interesting series of magnetic phase transitions and magnetic
structures that are observed in this material.

\section{Zero-Field Magnetic Structures}

The nature of the magnetic phase transitions and temperature dependence of
the order parameters have already been given in our earlier work,\cite
{Grigereit,Huang} so here we will just briefly describe the evolution of the
magnetic scattering as a function of temperature while providing some
further experimental details. Fig.~4 shows the evolution of the magnetic
scattering at six different temperatures. A three-peak structure is already
evident for temperatures just above 8~K, indicating that long range order
has set in as the scattering intensity initially becomes observable. The
peak in the center belongs to the commensurate antiferromagnetic structure
as given in Fig. 1(b), while the satellites at q$_c$= 0.0543 \AA $^{-1}$ on
either side originate from the spiral magnetic structure [Fig. 1(c)]. We see
that initially the three peaks have about the same intensity---after taking
into account the combination of the magnetic form factor, f($\vec{\text{Q}}$%
), and the powder Lorentz factor, (sin$\theta $sin2$\theta $)$^{-1}$---and
increase in intensity at the same rate. However, below $\sim $5.3 K the
intensity of the antiferromagnetic peak begins to grow more rapidly while
the intensities of the spiral peaks plateau, and then drop sharply in
intensity, as shown in Fig. 5. This drop in intensity for the spiral peak
coincides with the reentrant superconducting transition temperature observed
in this material; the growth of the spiral amplitude forces the system back
to the normal conducting state, but when the spiral amplitude suddenly drops
then the superconducting state is restored. This behavior suggests that the
magnetic ordering is controlling this sequence of phase transitions. Thus we
believe the lock-in transition to the low temperature antiferromagnetic
state is being controlled by the magnetism, and this allows the
superconductivity to re-establish itself. Note in Fig. 1(c) that in the
spiral state there is a net uncompensated magnetization/exchange on the
(superconducting) Ni layers, and we believe this is what is causing the
magnetic/superconducting coupling and the reentrant behavior. The
antiferromagnetic structure (magnetic space group symmetry Pm$^{\prime }$mm$%
^{\prime }$), on the other hand, possess no such net coupling, and at low
temperatures this magnetic structure readily coexists with superconductivity.

The temperature dependence of the intensity for the spiral state, shown in
Fig. 5, indicates that there is some irreversibility on warming and cooling.
It also shows that even well below the reentrant transition there remains
some vestige of the spiral intensity, of the order of a few per cent.
However, Fig. 6 shows that the widths of these spiral peaks become quite
large at low T, indicating that these regions become quite small in size. We
therefore attribute this remaining scattering either to domain walls in the
crystalline particles or on their surface. Higher resolution studies show
that there is also a small intrinsic width for the commensurate
antiferromagnetic peaks, as the magnetic peaks are slightly broader than the
nuclear diffraction peak linewidths, and correspond to a magnetic domain
size of $\sim $2175~\AA\ at 5.3~K.\cite{Huang}

\section{Magnetic Field Dependence}

Field-dependent elastic neutron scattering measurements were performed in
order to further probe the interactions among the superconducting,
antiferromagnetic, and spiral magnetic states as functions of temperature
and field. We found that for the initial application of the field the
spiral, antiferromagnetic, and (00$\ell $)-type nuclear reflections all
increased in intensity, demonstrating that the powder crystallites were
reorienting in the field. Fig.~7 illustrates the initial observed field
dependence of the spiral and antiferromagnetic intensities, after cooling in
zero field to 5.3~K. The intensities first increase sharply to a peak near
0.25~T, indicating that the crystallites reorient in a relatively modest
field, while at higher fields both types of peaks decrease in intensity
gradually as the magnetic moments rotate into a ferromagnetic alignment. The
reorientation is caused by the {\em a-b} planes of the crystallites rotating
to become parallel with the externally applied magnetic field, showing that
the Ho$^{3+}$ magnetic moments strongly prefer the {\em a-b} plane. After
the initial field alignment, no further reorientation of the particles was
observed, and the sample remained aligned for all subsequent measurements.

The intrinsic field dependence of the intensities of the spiral and
antiferromagnetic peaks is shown in Fig. 8. These data were taken at our
lowest field-dependent temperature of 4.5~K, which is below the peak
temperature for the spiral state so that the commensurate (001)
antiferromagnetic peak is roughly three times as strong in intensity at zero
field. The behavior we show here is quite typical of that observed for all
higher temperatures; the data for 5.3 K have been previously presented.\cite
{Grigereit} We see that the initial response is for both components to start
slowly and monotonically decreasing with increasing field, and then they
decrease more rapidly above $\sim $0.3T. The positions of the peaks, on the
other hand, are essentially field independent as shown in Fig. 9. Here we
see that the positions of the spiral peaks might suggest a slight decrease
in q$_c$ at the higher fields, but it should be noted that this is in the
regime where the intensities are becoming small and the corresponding
statistical (and systematic) errors large. We have included the zero-field
temperature dependence of q$_c$ in the figure for comparison, where we note
that above the reentrant transition q$_c$ is also observed to be temperature
independent.

While the spiral and antiferromagnetic states both lose intensity with
increasing field strength due to the forced rotation of the spins to a
ferromagnetic alignment, the data in Fig.~8 show that they have different
functional dependencies, as well as clear evidence for hysteresis. In
particular, the spiral satellites are observed to have increased relative
intensities on returning the field towards zero as compared to increasing
field, while the commensurate peaks exhibit the opposite behavior. The
magnetic scattering is therefore being redistributed. The relative
scattering strengths are shown Fig.~10, where we plot the ratio of the
spiral to antiferromagnetic intensities at this temperature as a function of
field. We see that on increasing field the ratio has a maximum at $\sim $%
0.5~T, which maximum is a consequence of the antiferromagnetic intensity
decreasing more rapidly with field than the spiral intensity. After ramping
to high field, we see that we have a much larger peak (at $\sim $0.4~T) on
returning towards zero field, and indeed there is a peak in the spiral
intensity itself as shown in Fig. 8. Note that the competition and
hysteresis are only evident at non-zero field values; the scattering
intensity (essentially) returns to its initial value at zero field, as shown
in Fig.~8. Thus the application of modest magnetic fields clearly favors the
spiral state over the antiferromagnetic state; on increasing the field some
of the antiferromagnetic phase appears to transform to the spiral phase,
while on decreasing from high fields the spiral state preferentially forms
and resists transforming to the antiferromagnetic/superconducting phase. We
note that this field-dependent hysteresis is identical in origin to the
temperature-dependent hysteresis we observe in zero field as shown in Fig.
5. Here we see that on cooling from above T$_C$ the spiral state
preferentially forms, while on warming from low temperatures the
antiferromagnetic/superconducting state is preferentially maintained.

Finally we consider the magnetic field-induced response at the (002) nuclear
reciprocal lattice position. In Fig.~11 the vertical axis is labelled as
magnetization ($\mu _B$), induced by the application of H; the magnetization
is directly related to the square root of the integrated intensity of the
magnetic Bragg peak. Based upon the crystal field results, at 5.3~K only the
ground state doublet will be significantly involved. Thus to a first
approximation the magnetization is given by Boltzmann statistics for a
single spin-1/2 ion,
\begin{equation}
M={\mu }\text{tanh}[\frac \mu {k_BT}(H-H_e)],
\end{equation}
where $\mu $ is the net magnetic moment in Bohr magnetons at the Ho$^{3+}$
site, and H$_e$ is the exchange field. This exchange field represents the
apparent reduction in strength of the external magnetic field at the
internal lattice sites due to the antiferromagnetic exchange interaction
between neighboring planes.

For very low fields, M$\sim \chi $H, and the curve should be nearly linear.
However, for all fields below H$_{c2}$ screening currents will reduce the
field inside the superconductor, reducing the induced moment, while
decreasing the field from above H$_{c2}$ to below H$_{{\text{c}}1}$ can
result in significant trapped flux rather than the ideal internal field of
zero. To avoid these problems the data were modelled over a constrained
field range, empirically determined to be above $\sim $0.7~T. Above this
value the single-ion model works relatively well.

The results of the fit shown in Fig.~11 yield a net induced magnetic moment
of $8.25\pm 0.2\mu _B$ at 7T. The effective exchange field was found to be $%
0.37\pm 0.06$~T. Although the moments and fields obtained for opposite field
sweep directions are within the error range of each other, the data
themselves indicate that there is a small amount of hysteresis, possibly due
to a decreased superconducting fraction from trapped flux on reducing field.

\section{Discussion}

The {\em c}-axis spiral state, the antiferromagnetic arrangement of
ferromagnetic {\em a-b} planes at low T, and the superconductivity have all
been observed to coexist in every neutron scattering study thus far. \cite
{Grigereit,Huang,GoldmanStassisAmes,TomyOakridge,Lynn} The coincidence of
the giant (reentrant) anomaly in H$_{\text{c}2}$ and the maximum in the
intensity of the spiral magnetic component demonstrates that the spiral
state is unfavorable to the formation of Cooper pairs which undergo
increased pair-breaking as the spiral component grows to a maximum. This
pair-breaking likely originates from the net exchange field/magnetization on
the Ni layers as the magnetization vectors on each {\em a-b} plane rotate
away from the totally compensated antiferromagnetic arrangement to form the
spiral. From this viewpoint, which is supported by the temperature and
field-dependent hysteresis effects observed, the spiral state forms
naturally and is the preferred magnetic structure at intermediate
temperatures and fields. At lower temperature the incommensurate spiral
state locks-in to the commensurate antiferromagnetic structure, and thereby
superconductivity returns and coexists. This is quite different from the
situation in the ferromagnetic superconductors ErRhB$_4$\cite
{Moncton77,Sinha82} and HoMo$_6$(S,Se)$_6$\cite{Lynn81,Lynn84} wherein the
oscillatory state is formed as a compromise between the superconductivity
and the ferromagnetism.

The related ErNi$_2$B$_2$C compound does not exhibit a {\em c}-axis spiral,
but instead orders as a transversely polarized spin density wave along the
{\em a}-axis.\cite{Zarestky,SinhaEr} Bulk magnetization measurements also
show no anomalous minimum in the upper critical field of ErNi$_2$B$_2$C and
no significant hysteresis in the magnetic order parameter; there is only a
small anomaly in H$_{c2}$ near T$_N$ which is typical for antiferromagnetic
superconductors.\cite{FischerMaple} A small {\em a}-axis modulation has been
observed above the reentrant transition in HoNi$_2$B$_2$C,\cite
{GoldmanStassisAmes,TomyOakridge,Lynn} but it is clear that the {\em c}-axis
spiral is the component that competes with the superconductivity while the
{\em a}-axis modulation (if it exists in the pure phase of HoNi$_2$B$_2$C)
is not strongly coupled to the superconducting state.

Recent band theory calculations\cite
{Mattheiss1,PickettSingh,Mattheiss2,Coehoorn} have indicated that an unusual
combination of states, resulting in a peak in the density of states (DOS) at
the Fermi level of the Ni(3d) conduction band, is responsible for the
elevated values of T$_{\text{c}}$ in the borocarbide series. The
calculations also show the compounds to be three-dimensional metals\cite
{PickettSingh} despite their layered structure, unlike the
quasi-two-dimensional high-T$_{\text{c}}$ cuprates. Evidence supporting the
peak in the DOS is seen in pressure studies\cite
{SchmidtBraun,GaoChu,CaoChu,Alleno}, specific heat experiments,\cite
{Carter,Movshovich} the dependence of T$_{\text{c}}$ upon substitution\cite
{Lai,Bud'ko} and alloying\cite{Gangopadhyay}, but is not supported by
photoemission measurements,\cite{Golden} perhaps due to exchange correlation
effects. More recently evidence has been seen for two-dimensional
ferromagnetic spin wave contributions to the specific heat,\cite
{Movshovich,Canfield} and the observation of a cascade of magnetic phase
transitions\cite{Canfield,Massalami} in the range 0~T$\leq $H$\leq $1.5~T.
The transitions were identified as being consistent with the formation of a
fan structure and its response to an applied magnetic field. All of these
observations are consistent with the magnetic structure reported in our
earlier papers\cite{Grigereit,Huang} as shown in Fig.~1.

\section{Acknowledgments}

We would like to thank L. C. Gupta and S. K. Sinha for helpful
conversations. Research at the University of Maryland is supported by the
NSF, DMR 93-02380.

\section{Figure Captions}

Fig. 1. HoNi$_2$B$_2$C (a) Crystal structure; (b) Commensurate
antiferromagnetic structure; (c) Spiral magnetic structure.

Fig. 2. Inelastic intensities at 6~K of crystal field transitions out of the
ground state at 11.28 and 16.00 meV. The ``extra'' scattering on the
high-energy side of the elastic paramagnetic peak near zero energy transfer
likely originates from the ground state spin dynamics.

Fig. 3. Temperature dependence of the crystal field transitions; the dashed
lines are the fits to Boltzmann statistics. The two upper transitions out of
the ground state decrease with ground state depopulation as the temperature
increases while the intermediate transition between them increases.

Fig. 4. Temperature dependence of the magnetic diffraction pattern of the
spiral satellite peaks on either side of the (001) commensurate
antiferromagnetic peak at $\text{Q}\cong 0.59\text{\AA }^{-1}$. The central
commensurate antiferromagnetic (001) peak and the adjacent spiral peaks
develop near 8~K as in (a) and increase in intensity with decreasing
temperature as shown in (b) and (c). Below $\sim $5.3 K the (001)
commensurate antiferromagnetic peak grows rapidly in intensity (d-f) and
dominates at low T.

Fig. 5. (a) Integrated intensity of the commensurate antiferromagnetic
reflection as a function of temperature, showing thermal hysteresis. The
small dots are the peak counts versus T. (b) Temperature dependence of the
spiral satellites, observed on warming and cooling.

Fig. 6. Observed widths of the spiral and antiferromagnetic peaks versus T.
The rapid broadening of the spiral magnetic peaks below $\sim $5~K suggests
that the remnant spiral scattering at low T is due to a surface effect or to
domain walls.

Fig. 7. Reorientation of the powder to place the a-b plane of each
crystallite parallel to the externally applied vertical magnetic field
during the initial field ramp to 3~T. Both the (001) commensurate and the
(001)$^{+}$ high-angle satellite and (001)$^{-}$ low-angle satellite peaks,
as well as the (00$\ell $) nuclear reflections, show increased intensities
upon returning to zero field.

Fig. 8. Intrinsic field dependence of the intensities of the
antiferromagnetic and spiral peaks, for a temperature of 4.5 K and the field
applied in the {\em a-b} plane. Note that there is considerable hysteresis,
with the spiral state being preferred as the superconducting state is
approached from high fields.

Fig. 9. Temperature dependence (top) and field dependence (bottom) of the
positions of the spiral and antiferromagnetic peaks. There is little if any
field dependence to the satellite positions, and above $\sim 5$K they are
temperature independent as well.

Fig. 10. Ratio of the spiral satellite intensity to the antiferromagnetic
intensity, as a function of magnetic field.

Fig. 11. Field-induced magnetic moment at the (002) nuclear reciprocal
lattice position; at fields above 0.7~T the two-level ground state
single-ion model (solid line) fits the data well.

\end{document}